\documentclass[10pt, twocolumns]{IEEEtran}
\usepackage{amsmath,graphicx,amssymb,amsthm,bbm}
\usepackage{hyperref}

\usepackage{tikz}
\usetikzlibrary{decorations.markings}
\usepackage[caption=false,font=footnotesize]{subfig}
\usepackage{upgreek}
\usepackage{color}
\usepackage{soul}
\usepackage[boxruled]{algorithm2e}
\usepackage[english]{babel}
\SetAlFnt{\small}
\SetAlCapFnt{\small}
\SetAlCapNameFnt{\small}

\makeatletter
\renewcommand{\algocf@caption@boxruled}{%
  \hrule
  \hbox to \hsize{%
    \vrule\hskip-0.4pt
    \vbox{   
       \vskip\interspacetitleboxruled%
       \unhbox\algocf@capbox\hfill
       \vskip\interspacetitleboxruled
       }%
     \hskip-0.4pt\vrule%
   }\nointerlineskip%
}%
\makeatother

\title{An Active RBSE Framework to Generate Optimal Stimulus Sequences in a BCI for Spelling~\thanks{This work was supported by: NSF CNS-1136027, IIS1149570; NIH 2R01DC009834-06A1; NIDRR H133E140026. A complete package containing code and data associated with this document can be found online at the Northeastern University Library Digital Repository: \url{http://hdl.handle.net/2047/d20194049}.}}

\author{\IEEEauthorblockN{Mohammad~Moghadamfalahi\textsuperscript{1},~\IEEEmembership{Student~Member,~IEEE},\\
        Murat~Akcakaya\textsuperscript{2},~\IEEEmembership{Member,~IEEE},\\
        Hooman~Nezamfar\textsuperscript{1},~\IEEEmembership{Student~Member,~IEEE},\\
        Jamshid~Sourati\textsuperscript{1},~\IEEEmembership{Student~Member,~IEEE},\\
        and~Deniz~Erdogmus\textsuperscript{1},~\IEEEmembership{Senior~Member,~IEEE}}\\

\IEEEauthorblockA{\textsuperscript{1}Northeastern University\\
\textsuperscript{2}University of Pittsburgh\\
E-mails: \{moghamdam,nezamfar, sourati, erdogmus\}@ece.neu.edu, akcakaya@pitt.edu\\
Phone: +1-617-3733021}}
\begin{document}
\maketitle
\begin{abstract}
A class of brain computer interfaces (BCIs) employs noninvasive recordings of electroencephalography (EEG) signals to enable users with severe speech and motor impairments to interact with their environment and social network. For example, EEG based BCIs for typing popularly utilize event related potentials (ERPs) for inference. Presentation paradigm design in current ERP-based letter by letter typing BCIs typically query the user with an arbitrary subset characters. However, the typing accuracy and also typing speed can potentially be enhanced with more informed subset selection and flash assignment. In this manuscript, we introduce the active recursive Bayesian state estimation (active-RBSE) framework for inference and sequence optimization. Prior to presentation in each iteration, rather than showing a subset of randomly selected characters, the developed framework optimally selects a subset based on a query function. Selected queries are made adaptively specialized for users during each intent detection. Through a simulation-based study, we assess the effect of active-RBSE on the performance of a language-model assisted typing BCI in terms of typing speed and accuracy. To provide a baseline for comparison, we also utilize standard presentation paradigms namely, row and column matrix presentation paradigm and also random rapid serial visual presentation paradigms. The results show that utilization of active-RBSE can enhance the online performance of the system, both in terms of typing accuracy and speed.
\end{abstract}
\section{Introduction}
Noninvasive electroencephalography (EEG) based brain computer interfaces (BCIs) have shown promising results for establishing a safe alternative channel between the people with severe speech/muscle impairment and their environment~\cite{mur14}. BCIs can be used for different applications such as communication, environment control and wheelchair navigation~\cite{mur14,mog12}. For communication, letter by letter typing BCIs have been the subject of extensive research and development in the field~\cite{mur14,mog12,far88,sel03,orh12}. For instance, P300-matrix speller, which was first introduced by Donchin and Farewell, typically uses a letter by letter typing BCI~\cite{far88}. Even though matrix-based presentation scheme is very commonly used and various paradigms for this scheme were developed to improve typing speed and accuracy~\cite{sel06b,all03,jin11,tow10,tow12,jin15,li16,yeo14}, it has been shown the matrix-based presentation paradigms are highly gaze dependent and they cannot operate well for the population with covert attention\cite{tre10}. As a minimally gaze dependent alternative, rapid serial visual presentation (RSVP) paradigm has been offered in which all the symbols are presented in a pseudo-random order at a predefined location of the screen in a rapid serial manner~\cite{acq10,acq13,orh12,orh13,orh12b}. Generally, the performance comparison of different users on typing with row and column presentation (RCP), single character presentation (SCP), and RSVP paradigms, have demonstrated that at least for healthy population, the best presentation paradigm should be defined separately for each individual~\cite{mog15}.

We have developed a noninvasive EEG-based typing BCI which enables the user to choose among different matrix-based presentation paradigms and RSVP paradigm~\cite{mog15}. This system detects the user intent in a recursive Bayesian state estimation (RBSE) in which the state represents the user target. To improve the detection performance, we have incorporated a 6-gram language model that provides context priors to be probabilistically fused with the EEG likelihoods. Our earlier studies demonstrated a great benefit in using the language model both in terms of typing speed and accuracy~\cite{mog15,orh12}. In our system, a user is presented with a sequence of symbols and an event related potential in response to the desired character, in the recorded EEG, can indicate the user intent. To achieve a confident decision, our system might query the user with one or multiple sequences until a predefined confidence threshold is attained or an upper-bound on the number of sequences is reached ~\cite{orh12b,mog15,mog15b}.  In earlier versions of the system, each sequence contained the entire set of the symbols. However, experimentally, we observed that this method is very inefficient and adaptive subset selection methods are necessary to further improve the typing speed and accuracy.   

Here, we describe a sequence design strategy using the active learning concept. Active learning will be built upon RBSE for dynamic and time effective sequence optimization. The proposed query design and inference mechanism is denoted as active-RBSE framework. In this framework, we develop query strategies to actively choose a subset of stimuli for every presentation sequence. Specifically, in this manuscript, we propose to use a query function that employs the observed EEG evidence along with context information. According to this function, quires are adaptively specialized for the user, at every iteration, during each target detection step. We show that this query function is a modular and monotonic set function and accordingly the optimal solution for this function can be achieved with guaranteed convergence. We utilized real EEG data, to run Monte-Carlo simulation of our system when active-RBSE framework was utilized. The results show that, the proposed query function along with the time efficient optimization framework can improve the online typing performance in terms of speed and accuracy for both matrix-based presentation and RSVP paradigms. 
\section{General BCI system specifications}
In a previous study, we have shown that popular matrix-based presentation paradigms (namely 1. row and column presentation (RCP) paradigm, 2. single character presentation (SCP) paradigm), and rapid serial visual presentation (RSVP) paradigm in general provide comparable performances and the optimal presentation paradigm is different for each user~\cite{mog15}. Based on our previous results, we have decided to include all three presentation paradigms in the current design of our letter by letter BCI typing interface.  

We assume that the user intent is to select a character $x$ from a finite vocabulary set $\mathcal{V}$. In a typical typing scenario we can define $\mathcal{V}=\{A,~B,~,\cdots,~Z\}\cup \{<,~\_\}$ where $<$ represents the backspace for error correction and $\_$ is the space symbol. In a matrix presentation paradigm, one can propose to flash every arbitrary subset of vocabulary ${A}\subseteq\mathcal{V}$ in a ``\textit{trial}''. For example, in SCP the paradigm is such that, trials are singletons of the vocabulary set $A\in \mathcal{V},\,\,\, |{A}|=1$. Accordingly, we can record the EEG response to that trial $\mathbf{e}({A})$ for detection. A set of trials $\Phi=\{{A}_1,~{A}_2,~\cdots,{A}_{|\Phi|}\}$ which is presented to the user in a rapid serial manner is called a ``\textit{sequence}''. After every sequence the system attempts to infer the  target character intended by the user, but, due to low signal to noise ratio (SNR) of EEG usually the system need to query the user with more sequences to reach a decision with a predefined confidence level. On the other hand, in order to limit the time spent on typing each character the system will make a decision, regardless of confidence threshold, after a predefined number of sequences. In our system a set of sequences which lead to a decision is called an ``\textit{epoch}''. 
 \subsection{Probabilistic Graphical Model}
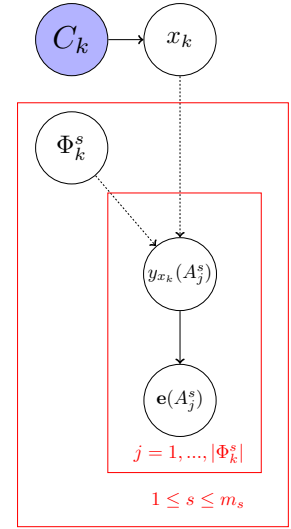
\begin{figure}[!ht]
        \begin{center}
            \scalebox{0.24}{
            \begin{tikzpicture}
\tikzstyle{main}=[circle, minimum size = 40mm, ultra thick, draw =black!100]
\tikzstyle{connect}=[-latex, thick]
\tikzset{
  big arrow/.style={
    decoration={markings,mark=at position 1 with {\arrow[scale=2,black]{>}}},
    postaction={decorate},
    shorten >=0.4pt}
    }
\draw (0,0) node[main,fill = blue!30] (context) {\scalebox{5}{$C_{k}$}};
\draw (6,0) node[main] (state) {\scalebox{4}{$x_k$}};
\draw (0,-6) node[main] (subsets) {\scalebox{4}{$\Phi_k^s$}};
\draw (6,-13) node[main] (labels) {\scalebox{3}{$y_{x_k}(A_j^s)$}};
\draw (6,-20) node[main] (features) {\scalebox{3}{$\mathbf{e}(A_j^s)$}};

\draw [black,ultra thick,big arrow] (context) -- (state);
\draw [black,dashed,ultra thick,big arrow] (state) -- (labels);
\draw [black,dashed,ultra thick,big arrow] (subsets) -- (labels);
\draw [black,ultra thick,big arrow] (labels) -- (features);

\draw[red,ultra thick] (2,-8.5) -- (10.5,-8.5) -- (10.5,-24) -- (2,-24) -- cycle;
\draw[red,ultra thick] (-3,-3.5) -- (12,-3.5) -- (12,-27) -- (-3,-27) -- cycle;

\draw (6.5,-22.25) node[below] {\color{red} \scalebox{3}{$j=1,...,|\Phi_k^s|$}};
\draw (7,-25) node[below] {\color{red} \scalebox{3}{$1\leq s\leq m_s$}};
\end{tikzpicture}}
        \end{center}
        \vspace{-5mm}
        \caption{Proposed probabilistic graphical model representing the 
        $k^{\rm th}$ epoch. Here, the dashed lines show a deterministic 
        relation, the solid lines define a probabilistic correspondence, and red rectangles represent conditional independence.}
        \label{fig:PGM}
    \end{figure}

The proposed probabilistic graphical model (PGM) illustrated in Fig~\ref{fig:PGM}. In this figure, $x_k$ represents the system state at epoch $k$, $\Phi_k^s$ is the set of trials selected (i.e. query set) to be flashed in the $s^{\rm th}$ sequence, $|\Phi_k^s|$ is the number of trials (i.e. the size of query set), $C_k$ is the context information, which in a typing scenario can be provided by a language model, and $e(A_j^s)$ is the EEG observation for the $j^{\rm th}$ trial in sequence $s$. In our model the label $y_{x_k}(A_j^s)$ for trial $A_j^s$ is a deterministic function of $x_k$, i.e.  $y_{x_k}(A_j^s):=\mathbbm{1}(x_k\in A_j^s)$, then $y_{x_k}(A_j^s)=1$ if $x_k\in A_j^s$ or $y_{x_k}(A_j^s)=0$ otherwise. Finally, the maximum number of sequences allowed in an epoch is denoted as $m_s$.
    
 Active-RBSE is built upon this graphical model and Within this framework, a generic AL and maximum a posteriori (MAP) inference loop will iterate by alternating between the following two steps:
    \begin{eqnarray}
        \text{Query:} 
            \widehat{\Phi}_k^{s+1}= 
            &\arg \max_{\Phi_k^{s+1}} g(\Phi_k^{s+1}) \hspace{5pt} \text{s.t.} \hspace{5pt} 
            \Phi_k^{s+1} \in \mathcal{F}_k \subseteq 2^{\mathcal{V}}
            \label{subeqn:querystrategy}\\
        \text{Inference:} \hat{x}_k = &
            \arg\max_{\text{x}}~P(x_k=\text{x}|\mathcal{E}^s, C, \{\Phi_k^i\}_{i=1}^s)
            \label{subeqn:mapinference}
    \end{eqnarray}
Here, $\Phi_k^{s+1}$ is a potential query set restricted to the set of feasible queries at time $k$, $\mathcal{F}_k$, which is a subset of all possible queries, $2^{\mathcal{V}}$, the power set of $\mathcal{V}$. The quality of a query from the perspective of AL is measured by the set function $g$. Note that we focus on MAP inference in this manuscript. Next, we describe how to obtain the MAP the inference and then in Section~\ref{sec:queryOptimization}, we will explain the query strategy. 

 \subsubsection{Recursive Bayesian State Estimation and Intent Inference}
Our systems employs the maximum a-posteriori (MAP) inference mechanism to detect the user intent. We use the well-known Bayes rule of "\textit{posterior }$\propto$ \textit{prior } $\times$ \textit{likelihood}" to estimate the posterior probability mass function (PMF) over the state space. For letter by letter typing, we use an n-gram language model (LM) to estimate the prior PMF over the character set, and the likelihood is obtained from EEG observations. Given the system state we assume, EEG measurements for sequences are independent from each others (see Fig~\ref{fig:PGM}). Based on this assumption we can update the posterior PMF recursively, after every sequence.   

The prior PMF from the LM along with the EEG observation likelihood is used in our system to calculate the posterior PMF over the vocabulary set after each sequence. Assume $1\leq s\leq m_s$ sequences have been shown to the user and define $\mathcal{E}^s=\{E^i\}_{i=1}^s$, where $E^i=\{e(A^i_j)|~j=1,~\cdots,~|{\Phi}_k^i|\}$. Similarly, take $Y_{x_k}^i=\{y_{x_k}(A_j^i)|~j=1,~\cdots,~|\Phi_k^i|\}$ then define $\mathcal{Y}_{x_k}^s=\{Y_{x_k}^i\}_{i=1}^s$. The MAP framework estimates the user intent by solving the following optimization problem:
    \begin{equation}
        \hat{x}_k=\arg\max_{\text{x}}~~P(x_k=\text{x}|\mathcal{E}^s,~ C, \{\Phi_k^i\}_{i=1}^s)
        \label{eq:MAPInference}
    \end{equation}
The posterior probability defined in (\ref{eq:MAPInference}) can be factorized in terms of likelihood and context prior using the assumptions imposed in Figure~\ref{fig:PGM}.
    \begin{equation}
    \small
        \begin{split}
           & P(x_k=\text{x}|\mathcal{E}^s,~ C, \{\Phi_k^i\}_{i=1}^s)=\frac{p(x_k=\text{x},~\mathcal{E}^s|~C,\{\Phi_k^i\}_{i=1}^s)}{p(\mathcal{E}^s|~C, \{\Phi_k^i\}_{i=1}^s)}\\
            &\propto p(\mathcal{E}^s|~x_k=\text{x}, \{\Phi_k^i\}_{i=1}^s)\cdot P(x_k=\text{x}|~C)
        \end{split}
        \label{eq:PosteriorEstimatior1}
    \end{equation}
But, for a given $x_k$ the $\mathcal{Y}_{x_k}^s$ for $\{\Phi_k^i\}_{i=1}^s$ is deterministically defined. Hence, according to the conditional independence of  $\mathcal{E}^s$ and context information defined in PGM we obtain;
    \begin{equation}
    \small
    \begin{split}
    &p(\mathcal{E}^s|~x_k=\text{x}; \{\Phi_k^i\}_{i=1}^s) = p(\mathcal{E}^s|~\mathcal{Y}_{x_k}^s; ~\{\Phi_k^i\}_{i=1}^s)=\\&\prod_{\substack{
            i=1,~\ldots,~s\\
            j=1,~\ldots,~|\Phi_k^i|}}^s {p(e(A_j^i)|y_{x_k}(A_j^i); ~\{\Phi_k^i\}_{i=1}^s)}
    \end{split}
            \label{eq:likelihood}
    \end{equation}
Then we can rewrite (\ref{eq:MAPInference}) as
    \begin{equation}
        \begin{split}
            &P(x_k=\text{x}|\mathcal{E}^s,~ C, \{\Phi_k^i\}_{i=1}^s)\propto \\& \prod_{\substack{
            i=1,~\ldots,~s\\
            j=1,~\ldots,~|\Phi_k^i|}}^s {p(e(A_j^i)|y_{x_k}(A_j^i);~\{\Phi_k^i\}_{i=1}^s)} \cdot P(x_k=\text{x}|~C)\propto \\
            &\prod_{\substack{
            i=1,~\ldots,~s\\
            \{j|~y_{x_k}(A_j^i)=1\}}} {\frac{p(e(A_j^i)|y_{x_k}(A_j^i)=1; ~\{\Phi_k^i\}_{i=1}^s)}{p(e(A_j^i)|y_{x_k}(A_j^i)=0; ~\{\Phi_k^i\}_{i=1}^s)}} \cdot P(x_k=\text{x}|~C)
        \end{split}
        \label{eq:posteriorWithLikelihoods}
    \end{equation}
In the next two subsections, we will describe how to obtain the context probability $P(x_k=\text{x}|~C)$ and the class conditional probabilities $p(e(A_j^i)|y_{x_k}(A_j^i)=1; ~\{\Phi_k^i\}_{i=1}^s)$ and $p(e(A_j^i)|y_{x_k}(A_j^i)=0; ~\{\Phi_k^i\}_{i=1}^s)$.
\subsubsection{n-gram Language Model for Context-based Prior Estimation}
In our system we utilize an n-gram language model (LM) to estimate the prior probabilities over the vocabulary set. The n-gram LM used in our system is essentially a Markov model of order $n-1$ which estimates a PMF over the state space for upcoming character, give the $n-1$ previously typed letters. 
 Let us define $C=\{x_l^*\}_{l=n-1,~\cdots,~1}$ as the order set of $n-1$ preceding typed characters where, $x^*_{l}$ represents the character at lag $l$. Then the probability of current character can be defined as follows:
    \begin{equation}
    \small
        P(x|C)=P(x|\{x_l^*\}_{l=n-1,~\cdots,~1})=\frac{P(x,~x_{n-1}^*,~\cdots,~x_1^*)}{P(x|~x_{n-1}^*,~\cdots,~x_1^*)}
        \label{eq:LM}
    \end{equation}
The LM used in our system, is trained on the NY Times portion of English Gigaword corpus~\cite{roa10}. 
This LM has shown to enhance the BCI typing performance in terms of typing accuracy and speed~\cite{orh12b,orh13}.

 \subsubsection{Class Conditional Distributions for EEG observations}
To estimate the class conditional distributions we collect labeled data in a "calibration session". Typically, during a calibration session the user is presented with $100$ sequences. Prior to presentation the user is asked to focus on a predefined target character during that sequence.

In our system, we record the signal from 16 EEG channels. To improve the SNR we apply a bandpass filter with the pass band of $[1.5,42]$ Hz and a notch filter at $60$ Hz to further attenuate the line noise. According to the pass band of the filter and data acquisition sampling rate ($256$ Hz), we down-sample the signal by a factor of $2$ to eliminate non-informative time samples while avoiding aliasing. The preprocessed EEG in a time window of $[0,500)$ms from the onset of each stimulus is assigned to that trial as it's EEG measurement at each channel. Subsequently, we eliminate directions with zero or negligible variances by applying principle component analysis (PCA) on these EEG measurements at each channel separately. Finally, we concatenate the measurements from every channel to form the EEG feature vector $\mathbf{f}_j^s$ for $j^{\rm th}$ trial at sequence $s$ in the $k^{\rm th}$ epoch.

EEG is assumed to be a Gaussian process~\cite{fau07,zho08,mog15}, hence, we could utilize quadratic discriminant analysis (QDA) to project the EEG feature vector onto a one dimensional space with minimum expected classification risk. But QDA requires a full rank class conditional covariance matrix estimation while for a typical setup of our system this is not feasible because the feature vector dimensionality is relatively higher than number of observations at each class. Instead, we utilize regularized discriminant analysis (RDA) which applies regularization and shrinkage on estimated class conditional covariance matrices and makes them invertible~\cite{RDA}. 

Assume $\mathbf{f}_i$ is an $m$ dimensional feature vector and $y_i\in \{0,1\}$ is the binary label for $\mathbf{f}_i$ then the maximum likelihood estimator for class conditional covariance matrix for class $k\in \{0,1\}$ is;
   \begin{equation}
        \begin{split}
            &\boldsymbol{\upmu}_k =\frac{1}{N_k}\sum_{i=1}^N
                {\mathbf{f}_i\delta(y_i,k)} \\
            &\mathbf{\Sigma}_k =\frac{1}{N_k}\sum_{i=1}^N
                {(\mathbf{f}_i-\boldsymbol{\upmu}_k)
                (\mathbf{f}_i-\boldsymbol{\upmu}_k)^T\delta(y_i,k)} 
        \end{split}
    \end{equation}
for which $\delta(.,.)$ represents the Kronecker-$\delta$ function and $N_k$ is the number of observations in class $k$. Then the regularization and shrinkage steps are applied as follows:
    \begin{equation}
        \begin{split}
            &\widehat{\mathbf{\Sigma}}_k(\lambda)=\frac{(1-\lambda)
                N_k\mathbf{\Sigma}_k+(\lambda)\sum_{k=0}^1{N_k\mathbf{\Sigma}_k}}
                {(1-\lambda) N_k+(\lambda)\sum_{k=0}^1{N_k}}  \\
            &\widehat{\mathbf{\Sigma}}_k(\lambda,\gamma)=
                (1-\gamma)\widehat{\mathbf{\Sigma}}_k(\lambda)+(\gamma) 
                \frac{1}{m}tr[\widehat{\mathbf{\Sigma}}_k(\lambda)]\mathbf{I}_m
        \end{split}
    \end{equation}
where $\lambda,\gamma\in [0,1]$ are the shrinkage and regularization parameters, $tr[\cdot]$ is the trace operator and $\mathbf{I}_m$ is an identity matrix of size $m\times m$. To optimize the values of $\lambda$ and $\gamma$, in our system we apply $10$-fold cross validation to maximize the area under the receiver operating characteristics (ROC) curve (AUC). In our system the "EEG evidence" for trial $A_j^i$ is computed as follows.
    \begin{equation}
        e(A_j^i)=\log\left(\frac{f_\mathcal{N}
        (\mathbf{f}_j^i;\boldsymbol{\upmu}_1,
        \widehat{\mathbf{\Sigma}}_1(\lambda,\gamma))}
        {f_\mathcal{N}(\mathbf{f}_j^i;\boldsymbol{\upmu}_0,
        \widehat{\mathbf{\Sigma}}_0(\lambda,\gamma))} \right)
    \end{equation}
Here, $\mathbf{f}_j^i$ is the EEG feature vector for trial $A_j^i$, $f_\mathcal{N}(\mathbf{f};\boldsymbol{\upmu},{\mathbf{\Sigma}})$ is the Gaussian probability density function with mean $\boldsymbol{\upmu}$ and covariance $\mathbf{\Sigma}$, and $\lambda,\gamma\in [0,1]$ are the shrinkage and regularization parameters. To optimize the values of $\lambda$ and $\gamma$, in our system we apply $10$-fold cross validation to maximize the area under the receiver operating characteristics (ROC) curve (AUC).
To obtain the class conditional probability distributions $p(e(A_j^i)|y_{x_k}(A_j^i)=1; ~\{\Phi_k^i\}_{i=1}^s)$ and $p(e(A_j^i)|y_{x_k}(A_j^i)=0; ~\{\Phi_k^i\}_{i=1}^s)$ over the EEG evidence we further apply kernel density estimation (KDE) with a Gaussian kernel of a bandwidth which is computed using Silverman rule of thumb~\cite{sil86}.
\section{Query set Optimization Using Active Learning}\label{sec:queryOptimization}
In this section, we propose that optimizing the query set using the prior information from the language model and the EEG signal in response to earlier sequences in a specific epoch can improve the typing performance for that epoch. As a claim to this proposition in our earlier studies we have shown that for RSVP paradigm, it is inefficient to present the whole or randomly selected subsets of vocabulary at every sequence~\cite{mog15b}. Here, we use active query selection inspired by the active learning concept to define a combinatorial optimization problem, which exploits previously acquired information to appoint the query set elements in a timely manner. 
\subsection{Objective Function}
In this section, we will consider a specific selection of $g(.)$ to use in active-RBSE framework, specifically in (\ref{subeqn:querystrategy}). Let us hypothesize that we know the actual user intent $x_k^*$ for current epoch (i.e. epoch $k$). Our objective is to define an optimal query set for sequence $s+1$ while $s$ sequences have been already queried for inference of $x_k^*$ but the confidence threshold is not attained yet. Then, before presenting the $(s+1)^{\rm th}$ sequence, we can obtain a prediction of posterior PMF for a given $\Phi_k^i$ as follows. 
We define a function $g:\mathcal{V},2^{2^{\mathcal{V}}}\to \mathbb{R}$ as:
\begin{equation}
    \begin{split}
        &g(\mathbf{x}, \Phi_k^{s+1})= P(x_k=\text{x}|\mathcal{E}^s,~ C, \{\Phi_k^i\}_{i=1}^{s+1}, ~x_k^*=\text{x})\\
        &=\int_{\tilde{E}^{s+1}}P(x_k=\text{x},\tilde{E}^{s+1}|\mathcal{E}^s,~ C,\{\Phi_k^i\}_{i=1}^{s+1}, ~x_k^*=\text{x}) d(\tilde{E}^{s+1})\\
        &=\mathbb{E}_{\tilde{E}^{s+1}|\Phi_k^{s+1},~x_k^*} [P(x_k=\text{x}|\tilde{E}^{s+1},~\mathcal{E}^s,~ C,\{\Phi_k^i\}_{i=1}^s, ~x_k^*)]\\
        &=\mathbb{E}_{\tilde{E}^{s+1}|\Phi_k^{s+1},~x_k^*} \left[\frac{\Pi^{s+1}(\text{x})\cdot p(\tilde{E}^{s+1}|x_k=\text{x},~\Phi_k^{s+1})}{\sum_{\text{v}\in \mathcal{V}} \Pi^{s+1}(\text{v})\cdot p(\tilde{E}^{s+1}|x_k=\text{v},~\Phi_k^{s+1})}\right]
   \end{split}
   \label{eq:SubObj}
  \end{equation}
 where $\Pi^{s+1}(\text{x})=P(x_k=\text{x}|\mathcal{E}^s,~ C, \{\Phi_k^i\}_{i=1}^s)$ represents the prior probability of $\text{x}\in \mathcal{V}$ before observing sequence $s+1$. Moving from the third line to the fourth of (\ref{eq:SubObj}), we use the following
\begin{equation}
\begin{split}
    &P(x_k=\text{x}|\tilde{E}^{s+1},~\mathcal{E}^s,~ C, \{\Phi_k^i\}_{i=1}^{s+1})= \\
    &\frac{\Pi^{s+1}(\text{x})\cdot p(\tilde{E}^{s+1}|x_k=\text{x},~\Phi_k^{s+1})}{\sum_{\text{v}\in \mathcal{V}} \Pi^{s+1}(\text{v})\cdot p(\tilde{E}^{s+1}|x_k=\text{v},~\Phi_k^{s+1})}
\end{split}
\end{equation}
where the denominator is the normalization constant.  
        
Note that, $g(\mathbf{x}, \Phi_k^{s+1})$ computes the posterior probability of hypothesized target for a particular $\Phi_k^{s+1}$ given previously observed EEG and context. But during the current epoch, $\mathbf{x}$ is yet to be estimated; and hence, we marginalize out the dependency on this unobserved random variable by computing the expected value of $g(\mathbf{x}, \Phi_k^{s+1})$ with respect to the most recent estimate of state space posterior PMF, $\Pi^{s+1}(\text{x})$.

Accordingly, The objective function for query set selection is then defined as follows.
    \begin{equation}
        \widehat{\Phi}_k^{s+1}=\arg\max_{\Phi_k^{s+1}} ~~\mathbb{E}_{\Pi^{s+1}(\text{x})} \left[ g(\mathbf{x}, \Phi_k^{s+1}) \right]
    \end{equation}
\subsection{Solution of the Optimization Problem}
In equation (\ref{eq:SubObj}), for fixed $\text{x}$ and $\Phi_k^{s+1}$, the argument inside the expectation is only a function of $\tilde{E}^{s+1}$. Accordingly, we define $\tilde{\boldsymbol{\sigma}}=\left[\tilde{\sigma}(A_1^{s+1}),~\cdots,~\tilde{\sigma}(A_{|\Phi_k^{s+1}|}^{s+1})\right]$, where\[ \tilde{\sigma}(A_j^i)=\frac{p(\tilde{e}(A_j^i)|1)}{p(\tilde{e}(A_j^i)|0)}\] and use \eqref{eq:posteriorWithLikelihoods}, to specify $\mathcal{F}:\mathbb{R}^{|\Phi_k^{s+1}|}\to \mathbb{R}$ such that,
    \begin{equation}
        \begin{split}
            \mathcal{F}(\tilde{\boldsymbol{\sigma}})=\frac{\Pi^{s+1}(\text{x})\cdot \prod_{
            \{j|~y_{\text{x}}(A_j^{s+1})=1\}} \tilde{\sigma}(A_j^{s+1}) }{\sum_{\text{v}\in \mathcal{V}} \Pi^{s+1}(\text{v})\cdot \prod_{
            \{j|~y_{\text{v}}(A_j^{s+1})=1\}} \tilde{\sigma}(A_j^{s+1})}
        \end{split}
        \label{eq:F}
    \end{equation}
To simplify the optimization --the reason for the simplification is described later in this section--, we approximate the $g(\mathbf{x}, \Phi_k^{s+1})$ using the  Taylor series expansion and the function defined in (\ref{eq:F}). 
    \begin{equation}
        \begin{split}
           &g(\mathbf{x}, \Phi_k^{s+1})=\mathbb{E}_{\tilde{E}^{s+1}|\Phi_k^{s+1},~x_k^*} \left[\mathcal{F}(\tilde{\boldsymbol{\sigma}})\right]=\\
           &\mathbb{E}_{\tilde{E}^{s+1}|\Phi_k^{s+1},~x_k^*} \left[ \mathcal{F}(\boldsymbol{\upmu}_{\boldsymbol{\sigma}})+\left(\tilde{\boldsymbol{\sigma}}-\boldsymbol{\upmu}_{\boldsymbol{\sigma}}\right)\cdot \nabla\mathcal{F}(\boldsymbol{\upmu}_{\boldsymbol{\sigma}})+\cdots \right]
        \end{split}
    \end{equation}
where $\boldsymbol{\upmu}_{\boldsymbol{\sigma}}=\mathbb{E}_{\tilde{E}^{s+1}|\Phi_k^{s+1},~x_k^*} [\tilde{\boldsymbol{\sigma}}]$. We propose to substitute the original function $g(\mathbf{x}, \Phi_k^{s+1})$ with its locally suboptimal linear approximation around the $\boldsymbol{\upmu}_{\boldsymbol{\sigma}}$. The proposed linear approximation is widely used in the field of signal processing~\cite{kay08}, especially when higher order central moments of the distribution are negligible. Typically, in our system, the estimated class conditional distributions are unimodal with small variance hence we assume this approximation is justifiable. Accordingly, we have:
    \begin{equation}
        \begin{split}
            &g(\mathbf{x}, \Phi_k^{s+1})\approx\hat{g}(\mathbf{x}, \Phi_k^{s+1})=\\
            &\mathcal{F}(\boldsymbol{\upmu}_{\boldsymbol{\sigma}})+\mathbb{E}_{\tilde{E}^{s+1}|\Phi_k^{s+1},~x_k^*} \left[\left(\tilde{\boldsymbol{\sigma}}-\boldsymbol{\upmu}_{\boldsymbol{\sigma}}\right)\right]\cdot \nabla\mathcal{F}(\boldsymbol{\upmu}_{\boldsymbol{\sigma}})=~\mathcal{F}(\boldsymbol{\upmu}_{\boldsymbol{\sigma}})
        \end{split}
        \label{eq:linearApprox}
    \end{equation}
Since we define  $\boldsymbol{\upmu}_{\boldsymbol{\sigma}}=\mathbb{E}_{\tilde{E}^{s+1}|\Phi_k^{s+1},~x_k^*} [\tilde{\boldsymbol{\sigma}}]$, the second term in equation~\eqref{eq:linearApprox} is equal to zero.
The next step, in our approach is to compute $\boldsymbol{\upmu}_{\boldsymbol{\sigma}}=\mathbb{E}_{\tilde{E}^{s+1}|\Phi_k^{s+1},~x_k^*} [\tilde{\boldsymbol{\sigma}}]$. Recall that according to the proposed graphical model in Figure~\ref{fig:PGM}, for a given $\mathcal{Y}_{x_k^*}^{s+1}$ the EEG evidence for different symbols, $A_j^{s+1}\in \Phi_k^{s+1} $, are independent from each other. Hence, $\tilde{\sigma}(A_i^{s+1})$ is independent from $\tilde{\sigma}(A_j^{s+1})$ $\forall i, j=1,\ldots, \left|\Phi_k^{s+1}\right|$ such that $i\neq j$. $\tilde{\sigma}(A_j^{s+1})$ is evaluated at samples from the following distributions:
\[ \left\{\begin{array}{ll} \tilde{e}(A^{s+1}_j)\sim p(e(.)|1),&\text{if}~~x_k^*\in A^{s+1}_j\\
\tilde{e}(A^{s+1}_j)\sim p(e(.)|0),&\text{if}~~x_k^*\not\in A^{s+1}_j
\end{array}\right.\]
Hence, by defining $\boldsymbol{\upmu}_{\boldsymbol{\sigma}}=[\hat{\sigma}(A_1^{s+1}),~\cdots,~\hat{\sigma}(A_{|\Phi_k^{s+1}|}^{s+1})]$, such that
\begin{equation}
\small
        \hat{\sigma}(A_j^{s+1})=\left\{\begin{array}{ll}
        \widehat{\sigma^+}=\mathbb{E}_{e(.)|1}
        \left[\frac{p\left(e(A_j^{s+1})|1\right)}
        {p\left(e(A_j^{s+1})|0\right)}\right],
        &\text{if}~ x_k^*\in A^{s+1}_j  \\\\
        \widehat{\sigma^-}=\mathbb{E}_{e(.)|0}\left[
        \frac{p\left(e(A_j^{s+1})|1\right)}
        {p\left(e(A_j^{s+1})|0\right)}\right],
         &\text{if}~  x_k^*\not\in A^{s+1}_j 
        \end{array}\right.
        \label{eq:pointEstims}
    \end{equation}
we can estimate  $ \hat{g}(\mathbf{x}, \Phi_k^{s+1})$ \footnote{Note that the approximation in \eqref{eq:linearApprox}, also corresponds to defining a point estimate of the EEG scores by calculating their mean value as computed in (\ref{eq:pointEstims}).} as follows:
    \begin{equation}
    \small
        \begin{split}
            \hat{g}(\mathbf{x}, \Phi_k^{s+1})=\frac{\Pi^{s+1}(\text{x})\cdot \widehat{\sigma^+}^{c^+_{\text{x,x}}(\Phi_k^{s+1})}}{\sum_{\text{v}\in \mathcal{V}}\Pi^{s+1}(\text{v})\cdot \widehat{\sigma^+}^{c^+_{\text{x,v}}(\Phi_k^{s+1})}\cdot \widehat{\sigma^-}^{c^-_{\text{x,v}}(\Phi_k^{s+1})}}
        \end{split}
        \label{eq:postpredict2}
    \end{equation}
in which, \[\begin{split}
            &c^+_{\text{x,v}}(\Phi_k^{s+1}) = \sum_{j=1}^{|\Phi_k^{s+1}|}{y_{\text{x}}(A_j^{s+1})\cdot y_{\text{v}}(A_j^{s+1})} ~~\text{and}\\
            &c^-_{\text{x,v}}(\Phi_k^{s+1}) = \sum_{j=1}^{|\Phi_k^{s+1}|}{(1-{y}_{\text{x}}(A_j^{s+1}))\cdot y_{\text{v}}(A_j^{s+1})}
        \end{split}\] where $y_{x}(A_j^{s+1}) \in \{0,1\}$.
We approximate $g(\mathbf{x},\Phi_k^{s+1})$ with $\hat{g}(\mathbf{x},\Phi_k^{s+1})$ and rewrite the optimization problem as follows:
    \begin{equation}
        \begin{split}
            \widehat{\Phi}_k^{s+1}&=\arg\max_{\Phi_k^{s+1}} ~~\mathbb{E}_{\Pi^{s+1}(\text{x})} \left[ \hat{g}(\mathbf{x},\Phi_k^{s+1}) \right]\\
            &=\arg\max_{\Phi_k^{s+1}} ~~\log\left(\mathbb{E}_{\Pi^{s+1}(\text{x})} \left[ \hat{g}(\mathbf{x},\Phi_k^{s+1}) \right]\right)
        \end{split}
        \label{eq:linearApprox2}
    \end{equation}
Here since logarithm is a monotonically increasing function, taking the logarithm of the cost function does not change the solution. Then, to solve the optimization problem defined in (\ref{eq:linearApprox2}), we define a lower-bound for the objective function using Jensen's inequality as follows:
    \begin{equation}
        \begin{split}
            &\log\left(\mathbb{E}_{\Pi^{s+1}(\text{x})} \left[ \hat{g}(\mathbf{x},\Phi_k^{s+1}) \right]\right)\geq \mathbb{E}_{\Pi^{s+1}(\text{x})} \left[ \log\left(\hat{g}(\mathbf{x},\Phi_k^{s+1})\right) \right]=\\
            & \mathbb{E}_{\Pi^{s+1}(\text{x})} \left[ \log\left(\frac{\Pi^{s+1}(\text{x})\cdot \widehat{\sigma^+}^{c^+_{\text{x,x}}(\Phi_k^{s+1})}}{\sum_{\text{v}\in \mathcal{V}}\Pi^{s+1}(\text{v})\cdot \widehat{\sigma^+}^{c^+_{\text{x,v}}(\Phi_k^{s+1})}\cdot \widehat{\sigma^-}^{c^-_{\text{x,v}}(\Phi_k^{s+1})}}\right)\right]\\
            & =\mathbb{E}_{\Pi^{s+1}(\text{x})} \left[\log(\Pi^{s+1}(\text{x}))+c^+_{\text{x,x}}(\Phi_k^{s+1})\log(\widehat{\sigma^+}) \right]-\\
            &\mathbb{E}_{\Pi^{s+1}(\text{x})}\left[  \log \left(\sum_{\text{v}\in \mathcal{V}}\Pi^{s+1}(\text{v})\cdot \widehat{\sigma^+}^{c^+_{\text{x,v}}(\Phi_k^{s+1})}\cdot \widehat{\sigma^-}^{c^-_{\text{x,v}}(\Phi_k^{s+1})}\right)\right]
        \end{split}
    \end{equation}
In our system, the class-conditional distributions are typically unimodal with small variances and different mean values, accordingly, we assume, $\widehat{\sigma^+}>1$ and $\widehat{\sigma^-}<1$. Furthermore, in our system we introduce an upper bound $|\Phi_k^{s+1}|\leq m_t$ to limit the number of trials in a sequence and provide a practical presentation paradigm. Then we get, 
    \[
        \begin{split}
            &\widehat{\sigma^+}^{c^+_{\text{x,v}}(\Phi_k^{s+1})}\leq (\widehat{\sigma^+})^{m_t}~~\text{and}~~\widehat{\sigma^-}^{c^-_{\text{x,v}}(\Phi_k^{s+1})}\leq \widehat{\sigma^-}^{0}=1.
        \end{split}
    \]
As a result $\widehat{\sigma^+}^{c^+_{\text{x,v}}(\Phi_k^{s+1})}\cdot \widehat{\sigma^-}^{c^-_{\text{x,v}}(\Phi_k^{s+1})}\leq (\widehat{\sigma^+})^{m_t}$. Hence,
    \begin{equation}
        \begin{split}
            & \mathbb{E}_{\Pi^{s+1}(\text{x})} \left[\log(\Pi^{s+1}(\text{x}))+c^+_{\text{x,x}}(\Phi_k^{s+1})\log(\widehat{\sigma^+}) \right]-\\
            &\mathbb{E}_{\Pi^{s+1}(\text{x})}\left[  \log \left(\sum_{\text{v}\in \mathcal{V}}\Pi^{s+1}(\text{v})\cdot \widehat{\sigma^+}^{c^+_{\text{x,v}}(\Phi_k^{s+1})}\cdot \widehat{\sigma^-}^{c^-_{\text{x,v}}(\Phi_k^{s+1})}\right)\right] \geq \\
            & \mathbb{E}_{\Pi^{s+1}(\text{x})} \left[\log(\Pi^{s+1}(\text{x}))+c^+_{\text{x,x}}(\Phi_k^{s+1})\log(\widehat{\sigma^+}) \right]-\\
            &\mathbb{E}_{\Pi^{s+1}(\text{x})}\left[  \log \left((\widehat{\sigma^+})^{m_t}\right)\right].
        \end{split}
        \label{eq:defLowerBound}
    \end{equation}
Note that in~\eqref{eq:defLowerBound}, the only term that is a function of $\Phi_k^{s+1}$ is $c^+_{\text{x,x}}(\Phi_k^{s+1})$; therefore, from~\eqref{eq:linearApprox2},    
    \begin{equation}
        \begin{split}
            \widehat{\Phi}_k^{s+1}&\approx\arg\max_{\Phi_k^{s+1}} ~~Q=\mathbb{E}_{\Pi^{s+1}(\text{x})} \left[c^+_{\text{x,x}}(\Phi_k^{s+1})\log(\widehat{\sigma^+}) \right]
            \label{eq:lowerB}
        \end{split}
    \end{equation}
\subsection{Combinatorial Optimization}
The approximated objective function defined in (\ref{eq:lowerB}) is a modular and monotonic set function $Q:2^{2^{\mathcal{V}}}\to \mathbb{R}$; therefore the optimization defined in \eqref{eq:lowerB} has guaranteed convergence properties~\cite{nem78}. Here, we prove that $Q$ is a monotone modular set function.\\

\textbf{Definition 1 .} (Discrete derivative~\cite{kra12})  
        
        \textit{ Assume a set function $f:2^{\mathcal{W}}\to \mathbb{R},~B\subseteq\mathcal{W},$ and $w\in \mathcal{W}$, then $\Delta_f(w|B):=f(B\cup\{w\})-f(B)$ is ``discrete derivative'' of $f$ at $B$ with respect to $w$.
        }\\
        
\textbf{Definition 2 .} (Modularity~\cite{kra12})  
        
        \textit{A function $f:2^{\mathcal{W}}\to \mathbb{R}$ is ``modular'' if for every $B_1\subseteq B_2\subseteq\mathcal{W}$ and $w\in \mathcal{W}\setminus B_2$, \[\Delta(w|B_1)=\Delta(w|B_2)\] or equivalently \[f(B_1\cap B_2)+f(B_1\cup B_2)=f(B_1)+f(B_2).\]
        }

\textbf{Lemma 1 .}  
        
        \textit{Take $\mathcal{W}=2^{\mathcal{V}}$, then the function $Q:2^{\mathcal{W}}\to \mathbb{R}$ as define in \eqref{eq:lowerB} is a modular set function.
        }\\

\small
\begin{proof}
Assume $\Phi_1\subseteq\Phi_2\subseteq 2^{\mathcal{V}}$ and $A\in 2^{\mathcal{V}}\setminus \Phi_2$, then 
    \[\begin{split}
        &\Delta_Q(A|\Phi_1)=\\
        &\mathbb{E}_{\Pi^{s+1}(\text{x})} \left[c^+_{\text{x,x}}(\Phi_1\cup \{A\})\log(\widehat{\sigma^+}) \right]-\mathbb{E}_{\Pi^{s+1}(\text{x})} \left[c^+_{\text{x,x}}(\Phi_1)\log(\widehat{\sigma^+}) \right]=\\
        & \mathbb{E}_{\Pi^{s+1}(\text{x})}\left[c^+_{\text{x,x}}(\Phi_1\cup \{A\})\log(\widehat{\sigma^+})-c^+_{\text{x,x}}(\Phi_1)\log(\widehat{\sigma^+})\right]
    \end{split}\]
Since $A\notin\Phi_1$, we use the definition of $c^+_{\text{x,x}}(.)$ to write
    \[\begin{split}
        &c^+_{\text{x,x}}(\Phi_1\cup \{A\})=c^+_{\text{x,x}}(\Phi_1)+c^+_{\text{x,x}}(\{A\})\Rightarrow \Delta_Q(A|\Phi_1)=\\
        &\mathbb{E}_{\Pi^{s+1}(\text{x})}\left[(c^+_{\text{x,x}}(\Phi_1)+c^+_{\text{x,x}}(\{A\}))\log(\widehat{\sigma^+})-c^+_{\text{x,x}}(\Phi_1)\log(\widehat{\sigma^+})\right]=\\
        & \mathbb{E}_{\Pi^{s+1}(\text{x})}\left[c^+_{\text{x,x}}(\{A\})\log(\widehat{\sigma^+})\right]
    \end{split}\]
Similarly as $A\notin\Phi_2$, we have 
    \[\begin{split}
        &\Delta_Q(A|\Phi_2)=\mathbb{E}_{\Pi^{s+1}(\text{x})}\left[c^+_{\text{x,x}}(\{A\})\log(\widehat{\sigma^+})\right]\Rightarrow\\
        &\Delta_Q(A|\Phi_1)=\Delta_Q(A|\Phi_2)
    \end{split}\]
\end{proof}

\normalsize
\textbf{Definition 3 .} (Monotonicity~\cite{kra12})

\textit{A set function $f:2^{\mathcal{W}}\to \mathbb{R}$ is ``monotone'' if for every $B_1\subseteq B_2\subseteq\mathcal{W}$, we get $f(B_1)\leq f(B_2)$.}\\

\textbf{Lemma 2 .}  
        
        \textit{Take $\mathcal{W}=2^{\mathcal{V}}$, then the function $Q:2^{\mathcal{W}}\to \mathbb{R}$ as define in \eqref{eq:lowerB} is a monotone set function.
        }\\

\small
\begin{proof}
Assume $\Phi_1\subseteq\Phi_2\subseteq 2^{\mathcal{V}}$, and define $\Phi_3=\Phi_2\setminus\Phi_1$, then  $\Phi_3\cup\Phi_1=\Phi_2$ and we can write:
    \[\begin{split}
        Q(\Phi_2)=\mathbb{E}_{\Pi^{s+1}(\text{x})} \left[c^+_{\text{x,x}}(\Phi_1\cup \Phi_3)\log(\widehat{\sigma^+}) \right]
    \end{split}\]
Moreover, $\Phi_3\cap\Phi_1=\varnothing$ then according to the definition of $c^+_{\text{x,x}}(.)$ we have
    \[\begin{split}
         &Q(\Phi_2)=\mathbb{E}_{\Pi^{s+1}(\text{x})} \left[(c^+_{\text{x,x}}(\Phi_1)+ c^+_{\text{x,x}}(\Phi_3))\log(\widehat{\sigma^+}) \right]=\\
         &\mathbb{E}_{\Pi^{s+1}(\text{x})} \left[c^+_{\text{x,x}}(\Phi_1)\log(\widehat{\sigma^+}) \right]+\mathbb{E}_{\Pi^{s+1}(\text{x})} \left[ c^+_{\text{x,x}}(\Phi_3)\log(\widehat{\sigma^+}) \right]=\\
         &Q(\Phi_1)+\mathbb{E}_{\Pi^{s+1}(\text{x})} \left[ c^+_{\text{x,x}}(\Phi_3)\log(\widehat{\sigma^+}) \right]
    \end{split}\]
Based on our assumption, $\widehat{\sigma^+}\geq 1\Rightarrow \log(\widehat{\sigma^+})\geq 0$. Also due to definition, $c^+_{\text{x,x}}(.)\geq 0$. Hence 
    \[\begin{split}
        &\mathbb{E}_{\Pi^{s+1}(\text{x})} \left[ c^+_{\text{x,x}}(\Phi_3)\log(\widehat{\sigma^+}) \right]\geq 0\Rightarrow\\
        &\mathbb{E}_{\Pi^{s+1}(\text{x})} \left[ c^+_{\text{x,x}}(\Phi_3)\log(\widehat{\sigma^+}) \right]+Q(\Phi_1)\geq 0+Q(\Phi_1)\Rightarrow\\
        &Q(\Phi_2)\geq Q(\Phi_1)
    \end{split}\]
\end{proof}
\normalsize
It has previously been shown that a deterministic greedy algorithm can provide a good approximation of the optimal solution for an $NP$-hard optimization problem with submodular and monotone objective functions within a guaranty bound from the optimal solution. Moreover, the greedy algorithm attains the optimal solution when the objective function is a modular monotone set functions~\cite{nem78}. 

For a fixed number of trials, $N_t$, in each sequence, \emph{deterministic greedy algorithm} is described in Algorithm~\ref{alg:Greedy}. This algorithm provides the global solution to the optimization problem defined in (\ref{eq:lowerB})
    \begin{algorithm}[ht!]
           \caption{Greedy algorithm for maximization of $Q$}
        \label{alg:Greedy}
            \textbf{Input:} The size of sequence set $N_t$.\\
            \textbf{Output:} Estimated sequence set $\widehat{\Phi}_k^{s+1}$.\\
            \noindent\rule{.5\columnwidth}{.5pt}\\
            \SetNlSkip{.5em}
            \DontPrintSemicolon
            \tcc{\footnotesize Initializations}
            \nl$\widehat{\Phi}_k^{s+1}\leftarrow \varnothing$\;
            \tcc{\footnotesize  Starting the Iterations}
            \nl\For{$i=1\to N_t$}{
            \tcc{\footnotesize Adding a the next optimal $A\in 
            2^\mathcal{V}\setminus \widehat{\Phi}_k^{s+1}$}
            \nl$\widehat{\Phi}_k^{s+1}\leftarrow \widehat{\Phi}_k^{s+1}\cup\{{\arg\max_{A}~\Delta_Q(A|\widehat{\Phi}_k^{s+1})}\}$ where $A\in
            2^\mathcal{V}\setminus \widehat{\Phi}_k^{s+1}$.\;
            }
            \nl\textbf{return} $\widehat{\Phi}_k^{s+1}$\;
    \end{algorithm}
\section{Experimental Resrults Results}\label{sec:expResults}
In this study, we used EEG data collected from 12 healthy individuals according to an IRB approved protocol for an earlier study~\cite{mog15}. The data was acquired from 16 EEG locations: Fp1, Fp2, F3, F4, Fz, Fc1, Fc2, Cz, P1, P2, C1, C2, Cp3, Cp4, P5 and P6 according to the International 10/20 configuration. To record the data we utilized g.USBamp bio-signal amplifier at the rate of 256Hz with active g.Butterfly electrodes. Each user performed three calibration sessions, one for each presentation paradigm i.e 1. RCP, 2.SCP and, 3. RSVP, at the presentation rate of $150$ms inter-trial-interval (ITI)\footnote{Here the ITI is referred to the time between the onsets of two consecutive trials.}. A calibration session of our system, consists of 100 sequences and prior to each sequence we ask the user to intent a  predefined character during that sequence. These data sets were used to obtain a the system parameters to be used in Monte-Carlo simulations of our BCI.

For each calibration data set we first estimate class conditional distributions over the target and non-target EEG evidences. We use the samples drawn from these distributions to perform $20$ Monte-Carlo simulations of the system. In each simulation, the system types missing words in $10$ different phrases with different difficulty levels \footnote{Lower levels consist of copying phrases that have letters which are assigned high probabilities by the language model. As the level increases, the language model probabilities become increasingly adversarial. Level $3$ is neutral on average.} from 1 (the easiest) to 5 (the most difficult). These phrases are selected uniformly across five difficulty levels. Here we report the results for simulated online performance of our system under proposed and baseline methods. One should note that the proposed method does not affect the calibration session as its goal is to enhance the system online performance by including the prior knowledge available in sequence design to avoid non-informative and irrelevant queries while obtaining most information about more probable choices. 

We report the results in terms of: (I) total typing duration (TTD) for typing $10$ phrases --which is inversely proportional to typing speed--, and (II) probability of phrase completion (PPC) --which is a measure of typing accuracy.

\subsection{RSVP Paradigm}\label{subsec:RSVPResults}
We used two sets of Monte-Carlo simulations (1) with random trial selection, and (2) with optimal query selection, to assess the effect active-RBSE on online system performance. The maximum number of sequences $m_t=8$ and number of trials within a sequence $k=14$, were selected based on our earlier experimental study~\cite{mog15b,mog15}. 
    \begin{figure}[ht!]
        \centering
        \includegraphics[width=.49\textwidth]{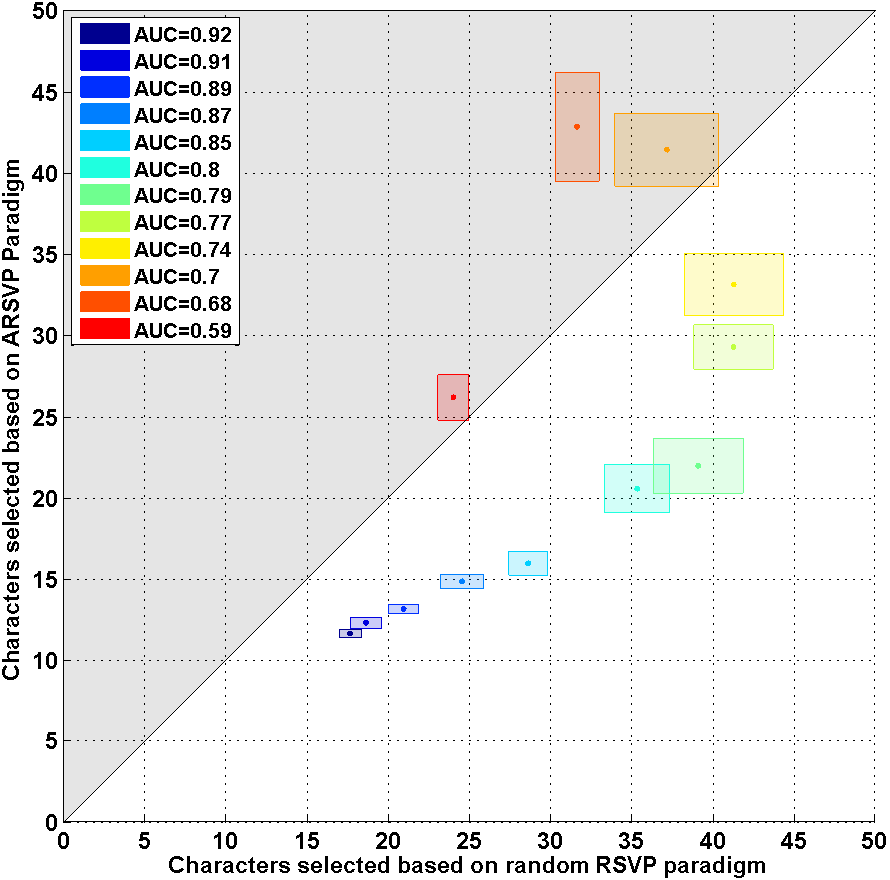}
        \vspace{-.2in}
        \caption{Scatter plot of average TTD in minutes from $20$ Monte-Carlo simulations. The horizontal axis shows the TTD when the sequences are selected randomly and the vertical axis represents the TTD for the optimal sequence selection. The height of the box around every point shows the standard deviation of TTD for random sequences and the width is the variance when sequences are optimized.}
        \label{fig:RSVPTotalTypingDuration}
    \end{figure} 
The TTD for active RSVP (ARSVP) vs. the random RSVP paradigm is presented in a scatter plot in Figure~\ref{fig:RSVPTotalTypingDuration}. In this figure, the horizontal axis represents TTD for random RSVP and the vertical axis show the TTD for typing with ARSVP. The width and the height of the box around each data point in the scatter plot shows the standard deviation of TTD from $20$ Monte-Carlo simulations for random and optimal sequences scenarios, respectively. 
According to this figure $9$ out of $12$ users could achieve a higher typing speed with optimal sequence selection. Wilcoxon signed-rank test result confirms a statistically significant improvement $(P < 0.03)$ in TTD among users.

\begin{figure}[ht!]
        \centering
        \includegraphics[width=.49\textwidth]{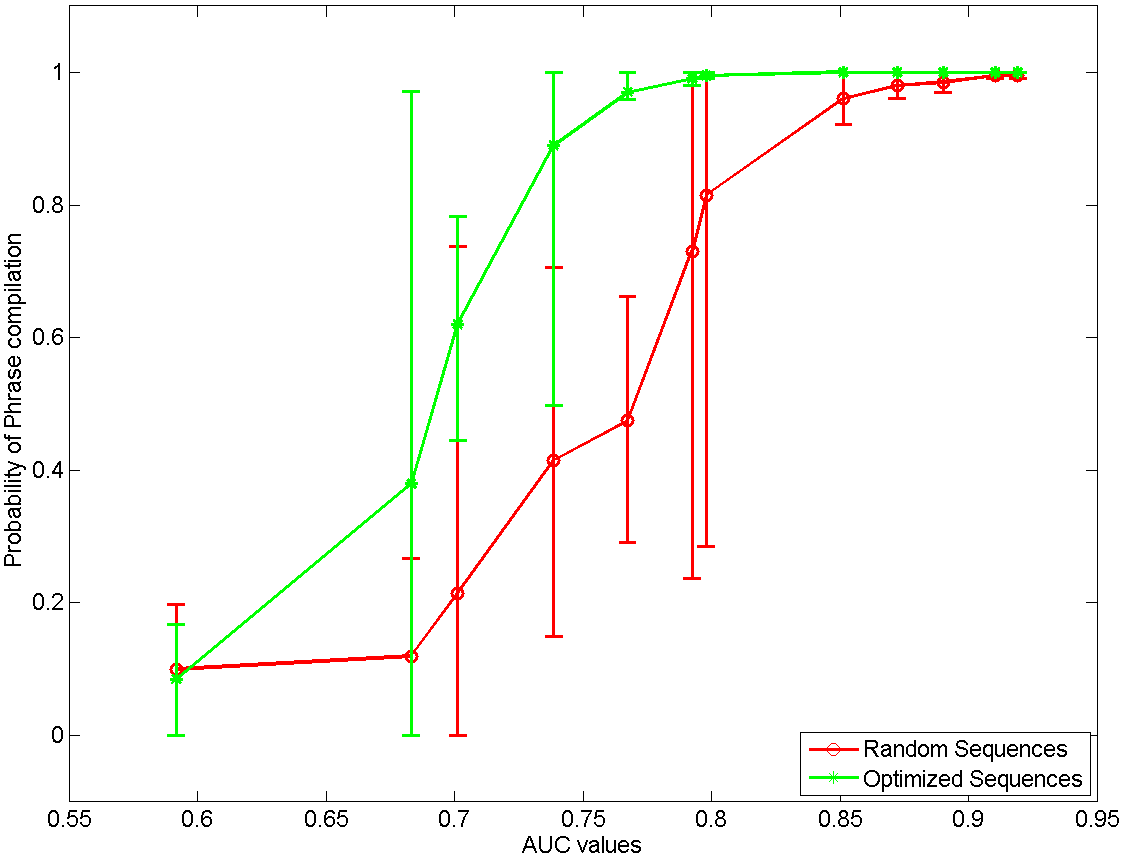}
         \vspace{-.1in}
        \caption{Average probability of phrase completion with 90\% confidence intervals for RSVP paradigm. The confidence interval is calculated by fitting a Beta distribution on PPC obtained from $20$ Monte-Carlo simulations.}
        \label{fig:RSVPPofPhraseComp}
    \end{figure}

In Monte-Carlo simulations, a phrase is assumed to be successfully completed, if the system types the correct phrase in a predefined duration with no more than five consecutive mistakes; otherwise, that phrase is assumed to be incomplete. Using this setup, the PPC obtained from the simulation sets are presented in Figure~\ref{fig:RSVPPofPhraseComp}. 
In this figure, the horizontal axis represents the AUC values for different participants. The green ``$*$'' points represent the averaged PPC from $20$ Monte-Carlo simulations and the error-bars represent the $90\%$ area under a beta distribution fitted to different PPCs obtained from ARSVP paradigm. Similarly, using the PPC values obtained from simulations with random RSVP paradigm, the mean PPC value, the red ``o'', and $90\%$ standard deviations, the red bars, are computed. 

The results in Figure~\ref{fig:RSVPPofPhraseComp} show that the optimal query strategy improves the typing accuracy. This effect is more clear for the AUCs$\in [0.7,0.9]$, and usually this range includes most of the user in healthy population. We measured the consistency of this improvement for all participants by performing Wilcoxon signed-rank test on average PPCs. The result demonstrates statistically significant improvement in PPC with $P < 0.003$.  
\subsection{Matrix-based Presentation Paradigm with Overlapping Trials}
Let us define a function $\mathbf{c}:2^{\mathcal{V}}\to {\{1,0\}}^{\mathcal{V}}$ such that $\mathbf{c}(A_i)=[\mathbbm{1}\{v_1\in A_i\},~\cdots,~\mathbbm{1}\{v_{|\mathcal{V}|}\in A_i\}]^T$. Accordingly, we define a $|\mathcal{V}|\times k$ code matrix $\mathbf{C}=[\mathbf{c}(A_1),~\cdots,~\mathbf{c}(A_k)]$. Then each row of the $\mathbf{C}$ matrix assigns a code word to each member of the vocabulary set which demonstrates its presence in trials of a sequence. 

RCP is the most widely used matrix-based presentation paradigm in which the trials have overlaps with $A_i\cap A_j\leq 1,~~\forall~i,j\in \{1,\cdots,|\Phi_k^{s+1}|\},~i\neq j$. If we define the number of rows and columns as $N_r$ and $N_c$ respectively, then the RCP paradigm offers unique codewords of length  $N_r$ + $N_c$ with two nonzero elements. In our experiments, we utilized a $4 \times 7$ background matrix of characters to efficiently distribute the $28$ symbols of our vocabulary set over the space available in wide-screen layout. In this layout $N_r=4$ and $N_c=7$ and the codewords length is $11$ for the RCP paradigm. 

In this study, we propose to define the search space such that each letter is uniquely identifiable from each sequence. When considering matrix presentation paradigms for ERP-based BCIs, one needs to consider some conditions for sequence set design, to satisfy the requirements for eliciting ERP. Here note that, in oppose to RSVP-based paradigms, matrix-based presentation paradigms can benefit from visual evoked potentials (VEPs) so we allowed for more frequent flashes of the same character in matrix based presentation paradigm~\cite{che13}. This can improve the typing speed by reducing the length of sequences. Accordingly, we propose to define the feasible set such that a unique code word is assigned to each symbol while, each symbol is presented with a probability of less than $0.5$ in each sequence. 

We set the codeword length $k=6$, to get enough codewords with at most $3$ non-zero elements. This setup offers ${6 \choose 3}+{6 \choose 2}+{6 \choose 1}=41$ unique code words to be assigned to each member of the vocabulary set. 
    \begin{figure}[ht!]
        \centering
        \includegraphics[width=.49\textwidth]{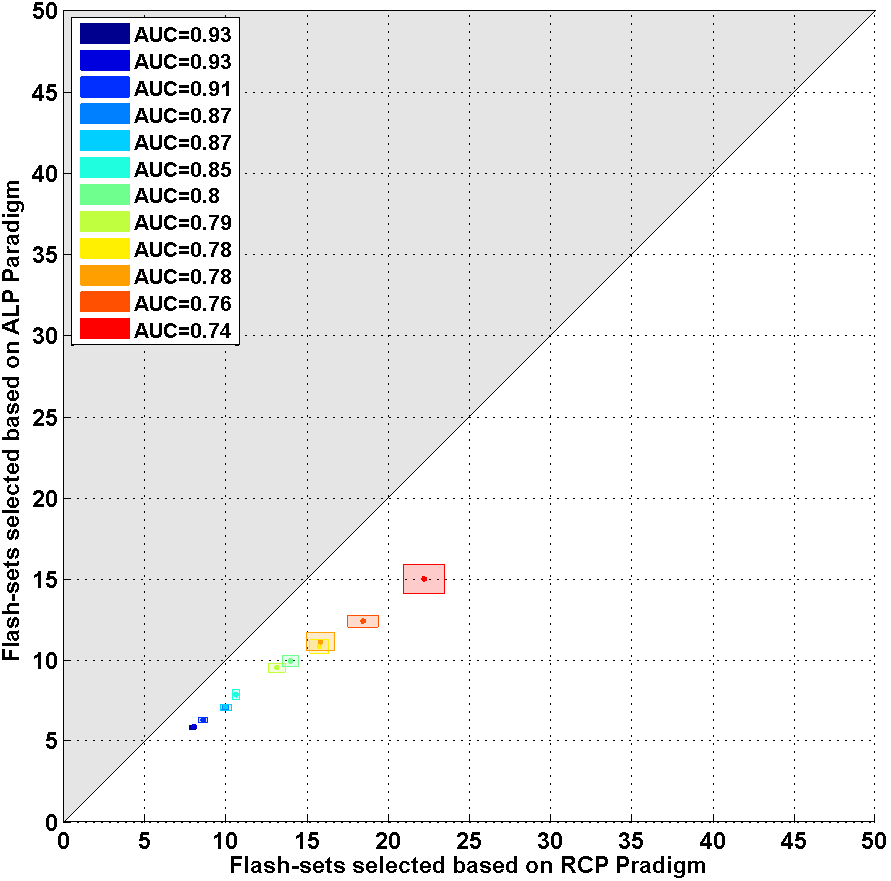}
        \vspace{-.2in}
        \caption{Scatter plot of total typing duration of $10$ phrases in terms of minutes. The horizontal axis shows the TTD when the sequences are selected based on RCP paradigm and the vertical axis represents the TTD for optimal sequence selection. The height of the box around every point shows the standard deviation of TTD for RCP paradigm and the width is the variance when sequences are optimized.}
        \label{fig:SCPTotalTypingDuration}
    \end{figure}  
The TTD comparison of RCP and actively learnt presentation (ALP) paradigm are presented in Figure~\ref{fig:SCPTotalTypingDuration}. The scatter plot suggests that the ALP can offer shorter TTD. The benefit of ALP is more visible for the lower AUCs which are demonstrated by the points concentrated in the center of the figure. Although the TTD improvement due to our proposed method is clear from this figure but, we performed Wilcoxon signed-rank test between average TTDs of ALP and RCP among all participants to demonstrate statistical evidence. Result confirms the statistical significant with $P<0.0005$. 
    \begin{figure}[ht!]
        \centering
        \includegraphics[width=.49\textwidth]{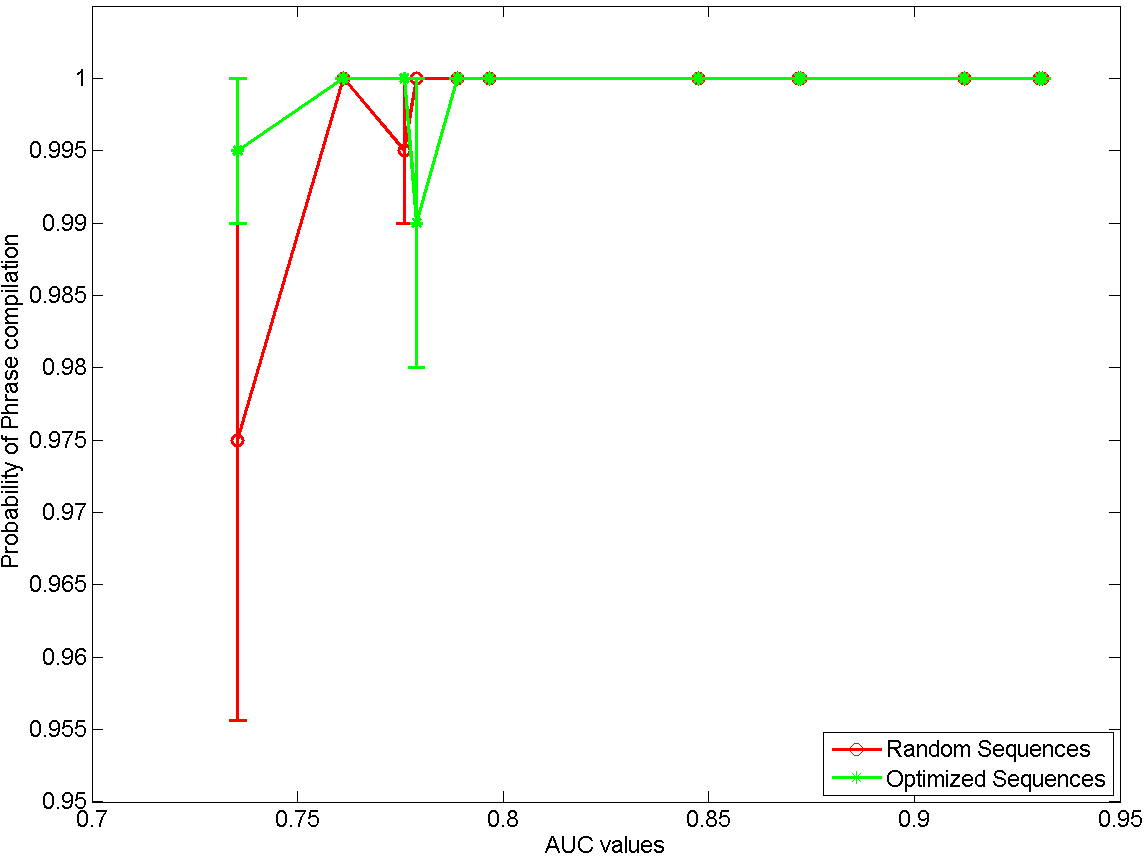}
         \vspace{-.1in}
        \caption{Average probability of phrase completion with 90\% confidence intervals.}
        \label{fig:SCPPofPhraseComp}
    \end{figure}
Figure~\ref{fig:SCPPofPhraseComp}, shows the effect of ALP on PPC in contrast to RCP. In RCP paradigm the participants AUCs are generally higher and consequently the PPC are above $\%95$ event without any sequence optimization. Thus our method did not offer any significant improvement for this case ($P>0.75$). As a conclusion, from Figures~\ref{fig:SCPTotalTypingDuration} and ~\ref{fig:SCPPofPhraseComp}, ALP can significantly reduce the TTD while preserving the PPC.
    
In a real typing scenario based on the assumptions, one can further propose to present a smaller subset of vocabulary while applying more restriction on the feasible space to prevent from repetition blindness or reduce the probability of each character in a sequence. 
\subsection{Matrix-based Presentation Paradigm with Single Character Trials}
In our earlier studies on optimal sequence length for RSVP paradigm, we have shown that the best typing performance can be achieved when not all letters but a subset of vocabulary is presented in each sequence~\cite{mog15b}. The matrix SCP paradigm is closely related to RSVP paradigm in the sense that each trial consists of a single letter and each letter will presented at most once in each sequence. Hence, we assume here that the best typing performance for SCP can be achieved with sequences of length $14$, similar to RSVP paradigm. Consequently, we assess the typing performance for optimizing the sequences of length $k=14$ where $|A_i|=1$, and compare it to typing performance obtained from the standard SCP paradigm in which all the vocabulary set would be flashed in every sequence. 

\begin{figure}[ht!]
        \centering
        \includegraphics[width=.49\textwidth]{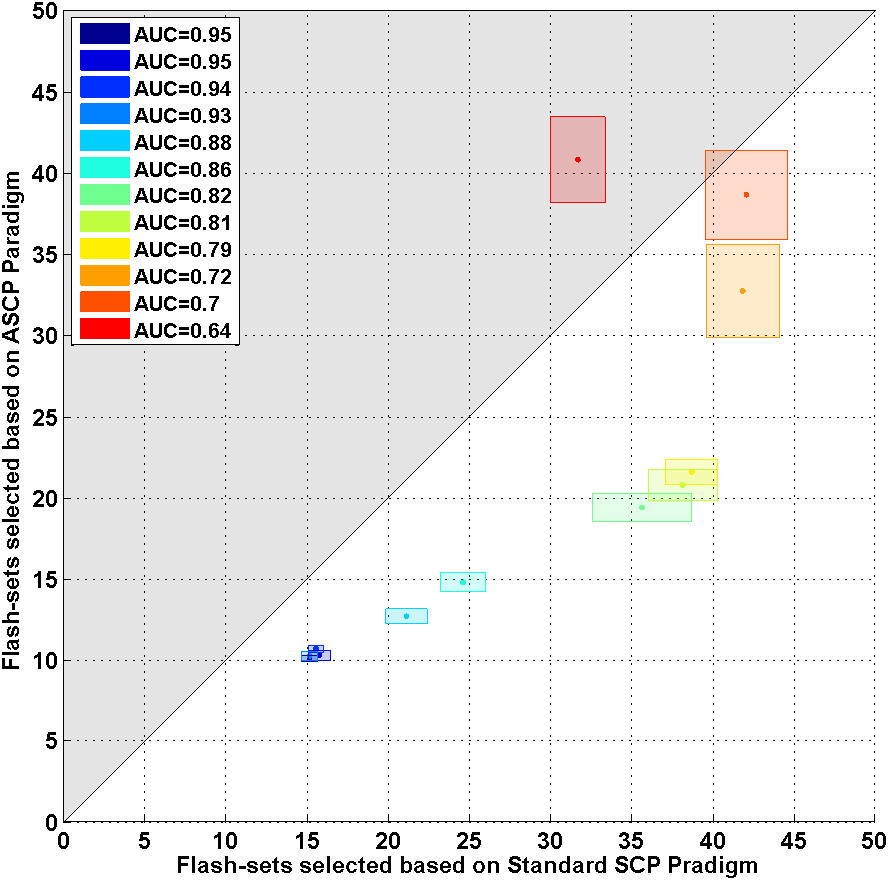}
        \vspace{-.2in}
        \caption{Scatter plot of TTD of $10$ phrases in terms of minutes. The horizontal axis shows the TTD when the sequences are selected based on SCP paradigm and the vertical axis represents the TTD for optimal sequence selection. The height of the box around every point shows the standard deviation of TTD for SCP paradigm and the width is the variance when sequences are optimized.}
        \label{fig:SingleTotalTypingDuration}
    \end{figure}

\begin{figure}[ht!]
        \centering
        \includegraphics[width=.49\textwidth]{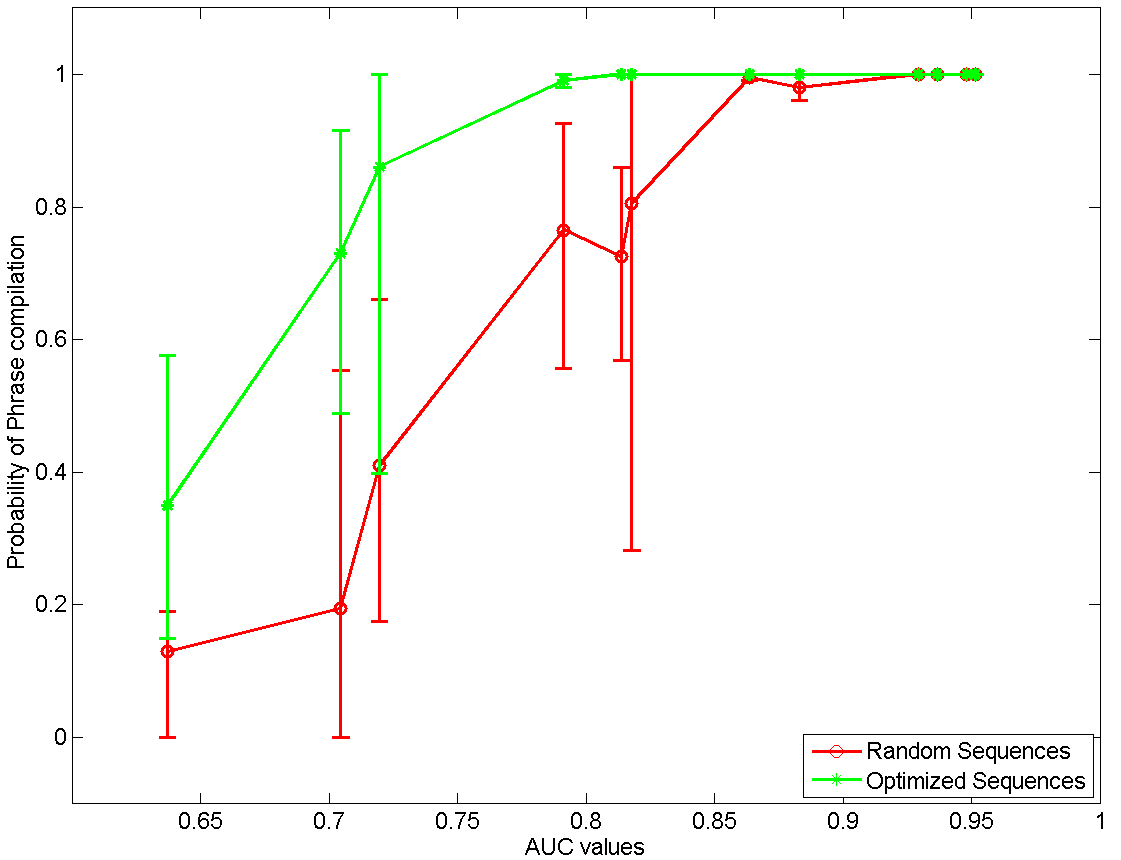}
         \vspace{-.1in}
        \caption{Average probability of phrase completion with 90\% confidence intervals.}
        \label{fig:SinglePofPhraseComp}
    \end{figure}

The results are summarized in Figures~\ref{fig:SingleTotalTypingDuration} \&~\ref{fig:SinglePofPhraseComp}. Figure~\ref{fig:SingleTotalTypingDuration} represents the scatter plot of TTD. The horizontal axis shows the TTD value for standard SCP paradigm and the vertical axis shows the TTD when the adaptive single character presentation (ASCP) paradigm is used for typing. This figure suggests that, the typing speed of a SCP paradigm can be significantly improved by optimally selecting smaller subset of characters $(P< 0.01)$.
More interestingly, the PPC comparison of ASCP and SCP as demonstrated in Figure~\ref{fig:SinglePofPhraseComp}, presents statistically significant improvements, with $P<0.008$, when an optimized subset of characters are used instead of full vocabulary set.
\section{Discussion and Future Work}
In this manuscript, we presented the active-RBSE framework which utilizes active-learning concept to optimize query sets in a noninvasive ERP-based BCIs. This framework is a mathematical establishment of our experimental observations in RSVP paradigm. Our goal was to demonstrate the usefulness of active-RBSE and providing a rough generalization to matrix-based presentation paradigms. 

For that, we used $36$ supervised data sets collected from $12$ healthy participants who utilized a language-model-assisted letter-by-letter typing interface with three different presentation paradigms of: RSVP, SCP, and RCP. Initial assessment of active-RBSE framework through Monte-Carlo simulations demonstrated that this framework offers a significant improvement in terms of typing speed and accuracy over well known existing presentation paradigms. We aim to experimentally validate these results in our future work. Also, we believe that this framework needs further improvements and analysis at least in three different categories. 

(I). The alternative bound on the original objective function used for solving optimizations problem does not factor-in overlaps between trials. This effect is mainly due to replacing the normalization factor with a fix upper-bound. In addition we need to provide the error introduced in the objective function value when the optimum point is estimated with this alternative bound.  

(II). The objective function presented in this manuscript was a particular choice based on intuition. This also corresponds to minimizing the expected value of R\'enyi entropy of order infinity obtained from predicted posterior PMFs. In future we can optimize the queries based on well established information theoretic concepts such as Shannon and R\'enyi entropy for exploration and/or exploitation. Moreover, we can relax the fixed sequence length assumption by allowing the algorithm to decide on query set size in every decision cycle. 

(III). This mathematical formulation allows us to consider human-in-the-loop effect both for decision making and query optimization by including physiological factors such as repetition blindness and effect of neighboring flashes which will be included in our future works.  

In our vision active-RBSE framework is not limited to EEG-based typing BCIs, we expect to utilize this framework not only under different sensor modalities and BCI applications but also for many other systems which utilize recursive querying in a sticky state estimation scenario.

\bibliographystyle{IEEEtran}
\bibliography{Ref.bbl}

\begin{thebibliography}{10}
\providecommand{\url}[1]{#1}
\csname url@samestyle\endcsname
\providecommand{\newblock}{\relax}
\providecommand{\bibinfo}[2]{#2}
\providecommand{\BIBentrySTDinterwordspacing}{\spaceskip=0pt\relax}
\providecommand{\BIBentryALTinterwordstretchfactor}{4}
\providecommand{\BIBentryALTinterwordspacing}{\spaceskip=\fontdimen2\font plus
\BIBentryALTinterwordstretchfactor\fontdimen3\font minus
  \fontdimen4\font\relax}
\providecommand{\BIBforeignlanguage}[2]{{%
\expandafter\ifx\csname l@#1\endcsname\relax
\typeout{** WARNING: IEEEtran.bst: No hyphenation pattern has been}%
\typeout{** loaded for the language `#1'. Using the pattern for}%
\typeout{** the default language instead.}%
\else
\language=\csname l@#1\endcsname
\fi
#2}}
\providecommand{\BIBdecl}{\relax}
\BIBdecl

\bibitem{mur14}
M.~Akcakaya, B.~Peters, M.~Moghadamfalahi, A.~Mooney, U.~Orhan, B.~Oken,
  D.~Erdogmus, and M.~Fried-Oken, ``{Noninvasive Brain Computer Interfaces for
  Augmentative and Alternative Communication},'' \emph{Biomedical Engineering,
  IEEE Reviews in}, vol.~7, no.~1, pp. 31--49, 2014.

\bibitem{mog12}
S.~Moghimi, A.~Kushki, A.~M. Guerguerian, and T.~Chau, ``{A Review of EEG-Based
  Brain-Computer Interfaces as Access Pathways for Individuals with Severe
  Disabilities},'' \emph{Assistive Technology: The Official Journal of RESNA},
  vol.~25, no.~2, pp. 99--110, 2012.

\bibitem{far88}
L.~Farwell and E.~Donchin, ``Talking off the top of your head: Toward a mental
  prosthesis utilizing event-related brain potentials,''
  \emph{Electroencephalography and clinical Neurophysiology}, vol.~70, pp.
  510--523, 1988.

\bibitem{sel03}
E.~Sellers, G.~Schalk, and E.~Donchin, ``{The P300 as a typing tool: tests of
  brain computer interface with an ALS patient},'' \emph{Psychophysiology},
  vol.~40, p.~77, 2003.

\bibitem{orh12}
U.~Orhan, K.~E. Hild, D.~Erdogmus, B.~Roark, B.~Oken, and M.~Fried-Oken,
  ``{RSVP keyboard: An EEG based typing interface},'' \emph{Acoustics, Speech
  and Signal Processing (ICASSP), 2012 IEEE International Conference on}, pp.
  645 -- 648, 2012.

\bibitem{sel06b}
E.~W. Sellers, D.~J. Krusienski, D.~J. McFarland, T.~M. Vaughan, and J.~R.
  Wolpaw, ``A {P}300 event-related potential brain--computer interface ({BCI}):
  the effects of matrix size and inter stimulus interval on performance,''
  \emph{Biological psychology}, vol.~73, no.~3, pp. 242--252, 2006.

\bibitem{all03}
B.~Z. Allison, J.~Pineda \emph{et~al.}, ``{ERPs evoked by different matrix
  sizes: implications for a brain computer interface (BCI) system},''
  \emph{Neural Systems and Rehabilitation Engineering, IEEE Transactions on},
  vol.~11, no.~2, pp. 110--113, 2003.

\bibitem{jin11}
J.~Jin, B.~Z. Allison, E.~W. Sellers, C.~Brunner, P.~Horki, X.~Wang, and
  C.~Neuper, ``{Optimized stimulus presentation patterns for an event-related
  potential EEG-based brain--computer interface},'' \emph{Medical \& biological
  engineering \& computing}, vol.~49, no.~2, pp. 181--191, 2011.

\bibitem{tow10}
G.~Townsend, B.~LaPallo, C.~Boulay, D.~Krusienski, G.~Frye, C.~Hauser,
  N.~Schwartz, T.~Vaughan, J.~Wolpaw, and E.~Sellers, ``{A novel P300-based
  brain--computer interface stimulus presentation paradigm: moving beyond rows
  and columns},'' \emph{Clinical Neurophysiology}, vol. 121, no.~7, pp.
  1109--1120, 2010.

\bibitem{tow12}
G.~Townsend, J.~Shanahan, D.~B. Ryan, and E.~W. Sellers, ``{A general P300
  brain--computer interface presentation paradigm based on performance guided
  constraints},'' \emph{Neuroscience letters}, vol. 531, no.~2, pp. 63--68,
  2012.

\bibitem{jin15}
J.~Jin, E.~W. Sellers, S.~Zhou, Y.~Zhang, X.~Wang, and A.~Cichocki, ``A p300
  brain--computer interface based on a modification of the mismatch negativity
  paradigm,'' \emph{International journal of neural systems}, vol.~25, no.~03,
  p. 1550011, 2015.

\bibitem{li16}
Y.~Li, J.~Pan, J.~Long, T.~Yu, F.~Wang, Z.~Yu, and W.~Wu, ``Multimodal bcis:
  target detection, multidimensional control, and awareness evaluation in
  patients with disorder of consciousness,'' \emph{Proceedings of the IEEE},
  vol. 104, no.~2, pp. 332--352, 2016.

\bibitem{yeo14}
S.-K. Yeom, S.~Fazli, K.-R. M{\"u}ller, and S.-W. Lee, ``An efficient erp-based
  brain-computer interface using random set presentation and face
  familiarity,'' \emph{PloS one}, vol.~9, no.~11, p. e111157, 2014.

\bibitem{tre10}
M.~S. Treder and B.~Blankertz, ``{Research (C) overt attention and visual
  speller design in an ERP-based brain-computer interface},'' \emph{Behavioral
  \& Brain Functions}, vol.~6, 2010.

\bibitem{acq10}
L.~Acqualagna, M.~S. Treder, M.~Schreuder, and B.~Blankertz, ``{A novel
  brain-computer interface based on the rapid serial visual presentation
  paradigm},'' in \emph{Engineering in Medicine and Biology Society (EMBC),
  2010 Annual International Conference of the IEEE}.\hskip 1em plus 0.5em minus
  0.4em\relax IEEE, 2010, pp. 2686--2689.

\bibitem{acq13}
L.~Acqualagna and B.~Blankertz, ``{Gaze-independent BCI-spelling using rapid
  serial visual presentation (RSVP)},'' \emph{Clinical Neurophysiology}, vol.
  124, no.~5, pp. 901--908, 2013.

\bibitem{orh13}
U.~Orhan, D.~Erdogmus, B.~Roark, B.~Oken, and M.~Fried-Oken, ``{Offline
  analysis of context contribution to ERP-based typing BCI performance},''
  \emph{Journal of neural engineering}, vol.~10, no.~6, p. 066003, 2013.

\bibitem{orh12b}
U.~Orhan, D.~Erdogmus, B.~Roark, B.~Oken, S.~Purwar, K.~E. Hild, A.~Fowler, and
  M.~Fried-Oken, ``{Improved accuracy using recursive Bayesian estimation based
  language model fusion in ERP-based BCI typing systems},'' in
  \emph{Engineering in Medicine and Biology Society (EMBC), 2012 Annual
  International Conference of the IEEE}.\hskip 1em plus 0.5em minus 0.4em\relax
  IEEE, 2012, pp. 2497--2500.

\bibitem{mog15}
M.~Moghadamfalahi, U.~Orhan, M.~Akcakaya, H.~Nezamfar, M.~Fried-Oken, and
  D.~Erdogmus, ``{Language-Model Assisted Brain Computer Interface for Typing:
  A Comparison of Matrix and Rapid Serial Visual Presentation},'' \emph{Neural
  Systems and Rehabilitation Engineering, IEEE Transactions on}, vol.~23,
  no.~5, pp. 910--920, Sept 2015.

\bibitem{mog15b}
M.~Moghadamfalahi, P.~Gonzalez-Navarro, M.~Akcakaya, U.~Orhan, and D.~Erdogmus,
  ``{The Effect of Limiting Trial Count in Context Aware BCIs: A Case Study
  with Language Model Assisted Spelling},'' in \emph{Foundations of Augmented
  Cognition}.\hskip 1em plus 0.5em minus 0.4em\relax Springer, 2015, pp.
  281--292.

\bibitem{roa10}
B.~Roark, J.~D. Villiers, C.~Gibbons, and M.~Fried-Oken, ``{Scanning methods
  and language modeling for binary switch typing},'' in \emph{Proceedings of
  the NAACL HLT 2010 Workshop on Speech and Language Processing for Assistive
  Technologies}.\hskip 1em plus 0.5em minus 0.4em\relax Association for
  Computational Linguistics, 2010, pp. 28--36.

\bibitem{fau07}
S.~Faul, G.~Gregorcic, G.~Boylan, W.~Marnane, G.~Lightbody, and S.~Connolly,
  ``Gaussian process modeling of eeg for the detection of neonatal seizures,''
  \emph{IEEE Transactions on Biomedical Engineering}, vol.~54, no.~12, pp.
  2151--2162, 2007.

\bibitem{zho08}
M.~Zhong, F.~Lotte, M.~Girolami, and A.~L{\'e}cuyer, ``Classifying eeg for
  brain computer interfaces using gaussian processes,'' \emph{Pattern
  Recognition Letters}, vol.~29, no.~3, pp. 354--359, 2008.

\bibitem{RDA}
J.~H. Friedman, ``Regularized discriminant analysis,'' \emph{Journal of the
  American statistical association}, vol.~84, no. 405, pp. 165--175, 1989.

\bibitem{sil86}
B.~W. Silverman, \emph{{Density estimation for statistics and data
  analysis}}.\hskip 1em plus 0.5em minus 0.4em\relax CRC press, 1986, vol.~26.

\bibitem{kay08}
S.~Kay, ``{Fundamentals of Statistical Signal Processing, Volume II: Detection
  Theory.}'' 2008.

\bibitem{nem78}
G.~L. Nemhauser, L.~A. Wolsey, and M.~L. Fisher, ``{An analysis of
  approximations for maximizing submodular set functions—I},''
  \emph{Mathematical Programming}, vol.~14, no.~1, pp. 265--294, 1978.

\bibitem{kra12}
A.~Krause and D.~Golovin, ``Submodular function maximization,''
  \emph{Tractability: Practical Approaches to Hard Problems}, vol.~3, p.~19,
  2012.

\bibitem{che13}
S.~Chennu, A.~Alsufyani, M.~Filetti, A.~M. Owen, and H.~Bowman, ``{The cost of
  space independence in P300-BCI spellers},'' \emph{Journal of neuroengineering
  and rehabilitation}, vol.~10, no.~82, pp. 1--13, 2013.

\end{thebibliography}
\end{document}